**Title: A Quantitative Image-Processing Framework for Eclipse Shadow Band Analysis**

**Authors:** Joseph P. Conti[1]*

**Affiliations:**

[1]Independent Researcher

*Corresponding author. Email: jconti@alum.mit.edu

**Abstract:** Eclipse shadow bands are transient intensity patterns that can appear on the ground near solar eclipse totality. This study presents a reusable image-processing framework for analyzing shadow-band video recordings collected with consumer-grade cameras. The framework quantifies band prominence, band orientation, band separation, band velocity and power spectral density from video recordings. Applied to two eclipse datasets, the method detected statistically significant shadow-band activity during eclipse windows that align with the scintillation theory for shadow bands. The findings also indicate that eclipse shadow bands can consist of simultaneous superimposed band modes with different orientations and velocities. This study demonstrates that consumer-grade cameras can support quantitative analysis of shadow bands and may support future observational and atmospheric studies.

## 1. Introduction

Eclipse shadow bands are faint lines that sometimes appear on the ground immediately before and after eclipse totality. These bands are typically aligned perpendicular to the path of the eclipse shadow, moving at a characteristic speed and sometimes changing orientation or intensity as totality approaches or recedes. Shadow bands are typically observed within the minutes prior to second contact, C2, when the Moon fully obscures the Sun, and the minutes immediately after third contact, C3, when the Sun begins to reappear from behind the Moon. These bands are best observed visually on a flat white surface when atmospheric conditions are calm and clear with little to no cloud cover. With observations dating back to the year 1842 *(1)*, the eclipse shadow band phenomenon is widely accepted to result from sunlight passing through small-scale refractive variations in Earth's atmosphere when the visible solar disk is reduced to a narrow crescent by the Moon *(2,3,4,5,6,7)*.

Previous studies have analyzed the power spectra and spatial correlations of photodiode measurements, confirming intensity fluctuations immediately before C2 and after C3 consistent with atmospheric scintillation acting on the narrow solar crescent. The evolution of band spacing has been measured from approximately 30 cm to 10 cm *(8)* and band velocities ranging from 1 to 50 m/s have been measured, consistent with typical atmospheric wind speeds *(9)*. Observations along different points of the eclipse center line during the same event show that shadow band orientation can vary between locations *(10),* further supporting an atmospheric origin. The influence of low-level atmospheric conditions has also been investigated using balloon-based measurements. In one study, photodiodes deployed on a balloon at approximately 25 km altitude detected a 4.5 Hz signal *(11)*. However, a follow-up study in 2024 using improved instrumentation did not reproduce this 4.5 Hz signal *(12)*. Modeling eclipse shadow bands using a wave optics approach has been used to study how different layers of the atmosphere contribute to shadow band effects *(13)*.

The main contributions of this study are as follows:

- A publicly available reusable and reproducible low-cost image-processing framework for quantifying multi-mode band prominence, band orientation, band separation, band velocity and power spectral density from ground-based consumer-grade videos.

- The use of an Orientation Distribution Function and Orientation-Prominence Metric for identifying dominant and secondary band modes.
- Quantitative detection of simultaneous orthogonal shadow band modes.
- Quantitative detection of rotating shadow bands.

## 2. Materials and Methods

### 2.1 Experimental Setup

A large white sheet approximately 1 meter square is placed on the ground with a digital camera on a stable stand pointing directly at the white sheet. For the North America dataset, an iPhone 13 camera with default settings was used to record video. The recorded video was then exported for analysis.

### 2.2 Method

**Flowchart**

Fig 1 depicts the overall image processing framework as a flowchart. The framework begins with loading a block of images from the experimental data and going through a pipeline of image processing steps.

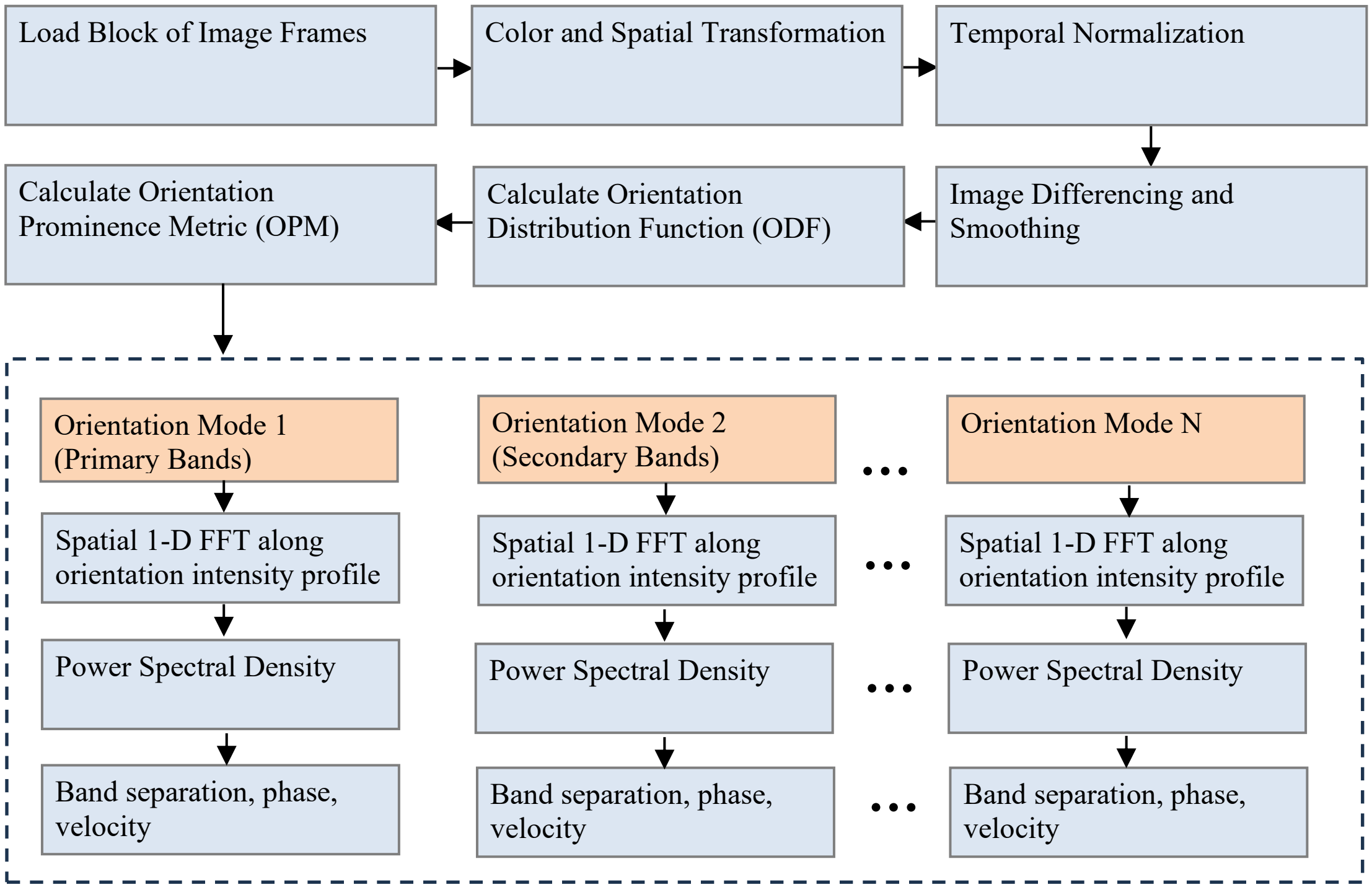

**Fig. 1. Flowchart for the processing framework.** The first part of the processing pipeline extracts significant orientation modes for shadow bands. The second part of the processing pipeline involves spatial spectral analysis to extract information like band separation and band velocity.

**Load Block of Image Frames**

The framework described in this study was conducted in MATLAB and is available for reuse (14). The framework begins by loading a sliding window block of images, $b$, from the video recording. The block size should be sufficient to avoid aliasing effects when extracting shadow bands. In this study, a block size of 3 frames was selected.

$$\mathrm{V} = (X_1, X_2, \dots, X_N)$$

$$\mathrm{B}_i = (X_i, X_{i+1}, \dots, X_{i+b-1}) \subset \mathrm{V}$$

$$t_i = \frac{i}{f_{video}}$$

$$f_{video} \geq 2 * f_{bands}$$

$$b = \frac{f_{video}}{2 * f_{bands}} > 1$$

where:

- V is a matrix representing the sequential set of image frames making up the video recording
- X is a matrix representing an image frame within the video
- $N$ is a scalar representing the total count of image frames within the video
- $i$ is a scalar representing an index of an image frame within a video
- $\mathrm{B}_i$ is a matrix representing the sliding block of sequential image frames starting at index $i$
- $b$ is a scalar representing the count of image frames in the sliding window block $\mathrm{B}_i$
- $f_{video}$ is a scalar representing the frame rate of the video [Hz]
- $t_i$ is a scalar representing the time stamp at index $i$
- $f_{bands}$ is a scalar representing the primary fundamental frequency of shadow bands [Hz]

**Color and Spatial Transformation**

Images are converted to grayscale *(16)*, transformed *(17)* from the original three-dimensional camera perspective to a two-dimensional view, and then transformed to be normal to the sun. These transformations ensure spatial consistency across different video recordings such that the working image for subsequent analysis is always orthographic view normal to the sun's direction. Bilinear interpolation is used for these transformations. The latter transformation requires measuring the orientation of the white sheet with respect to the sun. The MATLAB Image Processing Toolbox function FITGEOFORM2D was used to generate the transformation matrix from the pixel location of four corners representing a physical rectangle within the white sheet experimental setup. The MATLAB Image Processing Toolbox function IMWARP was then used to apply the subsequent transformations to the image.

$$\{R_i, G_i, B_i\} \in X_i$$

$$X_i^{gray} = .299 * R_i + .587 * G_i + .114 * B_i$$

$$B_i^{gray} = \left(X_i^{gray}, X_{i+1}^{gray}, \ldots, X_{i+b-1}^{gray}\right)$$

$$X_i^{xform} = WARP(X_i^{gray}, T^{camera}, T^{sun})$$

$$B_i^{xform} = \left(X_i^{xform}, X_{i+1}^{xform}, \ldots, X_{i+b-1}^{xform}\right)$$

where:

- R, G, B are matrices corresponding to the red, green and blue color channels for an image frame
- $X_i^{gray}$ is a matrix representing the image frame at index $i$ converted to grayscale
- $B_i^{gray}$ is a matrix representing the sliding block of sequential grayscale images starting at index $i$
- WARP() is a matrix valued function that warps the image for the supplied transformations
- $T^{camera}$ is a matrix representing the transformation from perspective view to orthographic view
- $T^{sun}$ is a matrix representing the transformation to be normal to the sun's orientation
- $X_i^{xform}$ is a matrix representing the transformed image
- $B_i^{xform}$ is a matrix representing the sliding block of sequential transformed images

**Temporal Normalization**

Temporal intensity normalization is applied to correct for ambient light variations that occur during the eclipse.

$$m_i = median(X_i^{xform})$$

$$m_i^{block} = median(\ B_i^{xform})$$

$$m_i^{correction} = m_i^{block} - m_i$$

$$X_i^{norm} = \ X_i^{xform} - m_i^{correction} \text{ (element wise scalar subtraction)}$$

$$X_i^{baseline} = \ \text{MEDIAN}(\ X_i^{norm}, X_{i+1}^{norm}, \ldots, X_{i+b-1}^{norm})$$

where:

- $median()$ is a scalar valued function that calculates the scalar median intensity from a matrix representing the image
- $m_i$ is a scalar representing the median intensity of an image
- $m_i^{block}$ is a scalar representing the median intensity over a block of images
- $m_i^{correction}$ is a scalar representing the normalization correction
- $X_i^{norm}$ is a matrix representing an image with normalized intensity
- MEDIAN() is a matrix valued function that calculates the pixel wise median from a set of images and outputs a matrix
- $X_i^{baseline}$ is a matrix representing element wise median over a block of images

**Image Differencing**

Image differencing is then calculated to extract shadow bands followed by Gaussian smoothing with a kernel size of 15 pixels *(18).* The scintillation index, the normalized variance of the measured intensity, is then calculated from the resulting smoothed image *(19).*

$$X_i^{diff} = X_i^{norm} - X_i^{baseline}$$

$$X_i^{smooth} = \text{SMOOTH}(X_i^{diff})$$

$$\sigma_i^2 = \frac{\langle {X_i^{smooth}}^2 \rangle - \langle X_i^{smooth} \rangle^2}{\langle X_i^{smooth} \rangle^2}$$

where:

- $X_i^{diff}$ is a matrix representing the result of image differencing
- $\text{SMOOTH}()$ is a matrix valued function that calculates the Gaussian smoothing for an image
- $X_i^{smooth}$ is a matrix representing the result of image smoothing
- $\sigma_i^2$ is a scalar representing the scintillation index for the image

### Orientation Distribution Function

Edge segmentation *(20,21)* with a threshold parameter of .1 and a standard deviation filter parameter of 30 is then performed to extract primary geometry. Line segments are extracted *(22,23)* and an Orientation Distribution Function (ODF) is calculated from the histogram of edge orientations with a bin size of 2 degrees. ODFs are commonly used in other scientific domains in fields including astronomy, biology, medical imaging, and quantum physics. The ODF is fitted with a two-term Gaussian model to measure the signal-to-noise (SNR) ratio of primary and secondary band modes. Having two terms ensures that any primary and secondary band orientation modes are measurable. A third, fourth, or higher-order term could also be used to detect additional shadow band modes but were not used in this analysis. An alternative approach to infer orientation would be to calculate a 2-D FFT, however, that approach was found to be less effective than the ODF-based method in this study.

$$\Theta_i = \{ \theta_1, \theta_2, \dots, \theta_M \} \in S_i$$

$$ODF_i(\theta) = \theta \in [0,180°)$$

$$ODF_{fit}(\theta, i) = a_{1,i} * \exp\left(-\left(\frac{\theta - \theta_{1,i}}{c_{1,i}}\right)^2\right) + a_{2,i} * \exp\left(-\left(\frac{\theta - \theta_{2,i}}{c_{2,i}}\right)^2\right) + d_i$$

where:

- $\theta$ is a scalar representing the orientation of each line segment
- $\Theta_i$ is the set of orientation values within $S_i$ at image frame index i
- $ODF_i(\theta)$ is the frequency distribution of orientations ranging from $0°$ $to$ $180°$ with a bin size of $2°$ at image frame index i
- $ODF_{fit}(\theta, i)$ is the fitted function for values of $h_k$ as a function of $\theta$ for the image at index $i$

### Orientation Prominence Metric

A dimensionless Orientation Prominence Metric (OPM) is calculated from the fitted $ODF_{fit}$ and corresponds to the signal-to-noise ratio (SNR) for the two terms of the Gaussian fit. The OPM serves as a metric for the intensity of band elongation across one or more modes and will be compared to the control groups to determine if band orientation is significant for each image frame.

$$OPM_{1,i} = \frac{a_{1,i}}{c_{1,i}}$$

$$OPM_{2,i} = \frac{a_{2,i}}{c_{2,i}}$$

where:

- $OPM_{1,i}$, $OPM_{2,i}$ are the primary and secondary orientation prominence metrics for the image frame at index $i$.

**Power Spectral Density**

A spatial power spectral density (PSD) is calculated by calculating the image intensity profile along a direction orthogonal to $\theta_{1,i}$ and then computing the spatial Fourier transform. To reduce noise, a PSD sample from the control period is subtracted.

$$X_{1,i}^{rotatedsmooth} = rotate(X_i^{smooth}, \theta_{1,i})$$

$$\mathrm{PSD}_{1,\mathrm{i}} = FFT\left(X_{1,i}^{rotatedsmooth}\right)$$

where:

- $X_{1,i}^{rotatedsmooth}$ is the rotated image frame at index $i$
- FFT() is a matrix value function that computes the 1-D FFT along an orientation
- $\mathrm{PSD}_{1,i}$ is the power along a range of spatial frequencies for a particular orientation

**Band Velocity**

Band velocity can be calculated from the change in fundamental phase across two sequential image frames.

$$(f_{1,\mathrm{i}}, \phi_{1,\mathrm{i}}, p_{1,\mathrm{i}}) = RIDGE(\mathrm{PSD}_{1,\mathrm{i}})$$

$$v_{1,\mathrm{i}} = (\phi_{1,\mathrm{i}} - \phi_{1,\mathrm{i}+1}) * f_{video}$$

where:

- RIDGE() is a scalar value function that computes the lowest frequency fundamental ridge from a PSD
- $f_{1,\mathrm{i}}$ is the spatial frequency for the fundamental ridge (Hz)
- $\phi_{1,\mathrm{i}}$ is the spatial phase for the fundamental ridge (°)
- $p_{1,\mathrm{i}}$ is the spatial power for the fundamental ridge (dB)
- $v_{1,\mathrm{i}}$ is the calculated band velocity (m/s)

**Statistical Rigor and Assumptions**

An empirical null distribution was constructed from all control-period frames collected between C2 and C3, when shadow bands are not expected to be present. For each frame, the Orientation Prominence Metric (OPM) was computed using the identical processing pipeline. Empirical p-values were calculated as the proportion of control-frame OPM values greater than or equal to the observed OPM. P-values were adjusted for multiple comparisons using the Benjamini-Hochberg false discovery rate procedure. Significant frames were defined as $q < 0.05$. The same methodology was applied to the PSD fundamental ridge power to determine whether ridge detections exceeded background levels. Because adjacent video frames are temporally correlated, the effective sample size may be smaller than the total number of analyzed frames.

## Results: 2019 South America Eclipse

The first experimental dataset analyzed in this study was recorded by Pauwels *(15)* on July 2, 2019, during the solar eclipse that traversed South America. The measurement site was located at −30.1768° S, −69.0984° W near Rodeo, Argentina, along the eclipse center line. A white sheet was mounted on a board and positioned so that its surface normal was oriented approximately toward the Sun. Totality occurred at approximately 4:39 pm local time, near sunset. Video recording began several minutes before C2 and continued until approximately 10 minutes after C3. Fig. 2 shows the image frame, image differencing, ODF and PSD at selected time points during the recording. Time is reported relative to totality, with $t = 0$ s defined as the midpoint between C2 and C3. At $t = 70$ s, approximately 2 s before C2, shadow bands are clearly visible, as indicated by a pronounced ODF peak at 78°, where $\theta = 0°$ corresponds to the positive x-axis of the transformed image and $\theta = 90°$ corresponds to the positive y-axis. At $t = 0$ s, when the Moon fully obscures the Sun, the ODF is flat, indicating no dominant shadow band orientation. At $t = -71$ s, approximately 2 s after C3, the ODF shows a strong peak near 30°. At $t = -102$ s, secondary shadow band modes appear at $\theta = 170°$, offset from the primary bands at $\theta = 40°$. A grid-like pattern is visually apparent, indicating the presence of multiple band orientation modes.

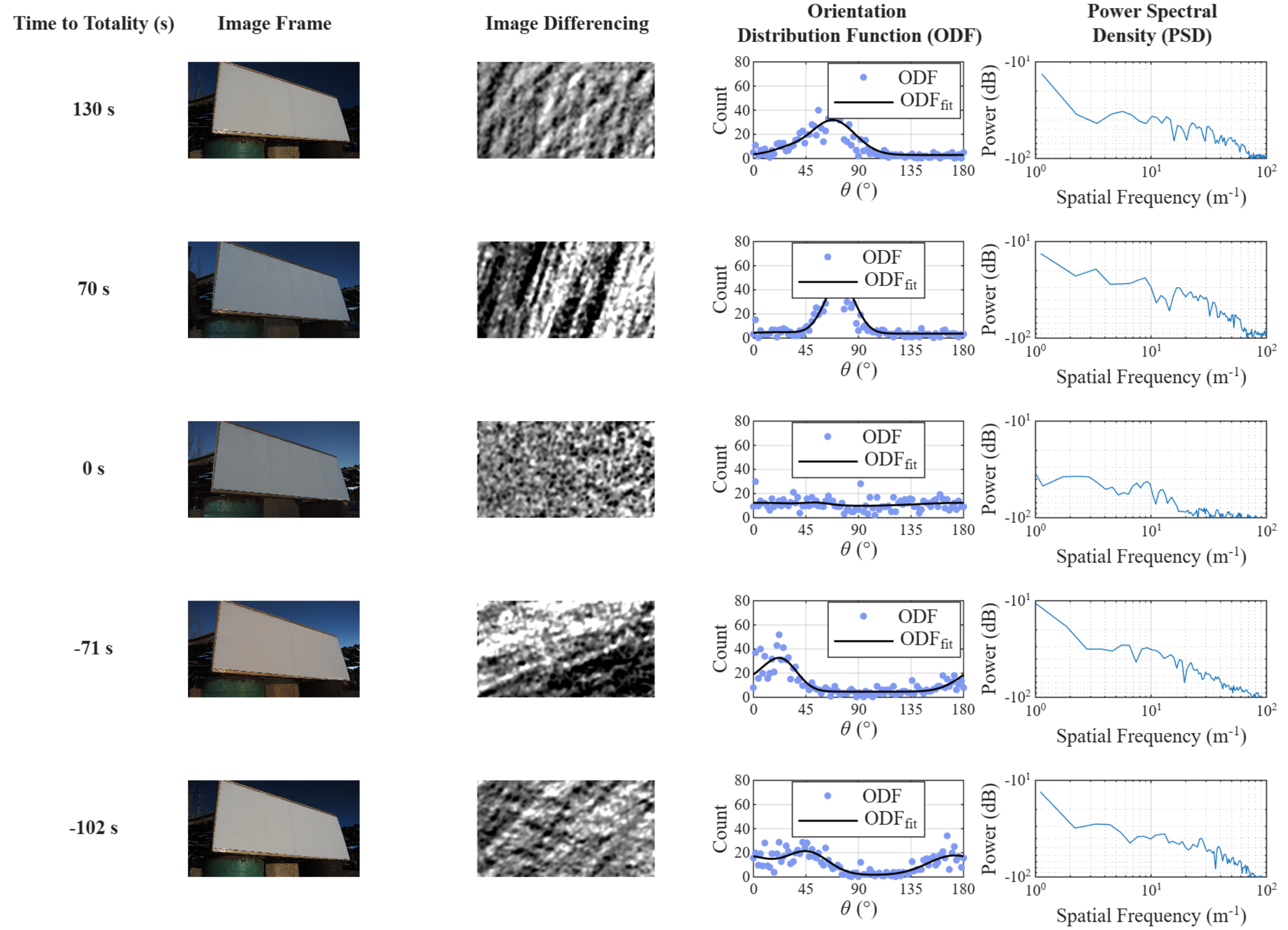

**Fig. 2. Image processing framework applied to the South American eclipse video data.** The orientation of the primary shadow bands rotates by approximately 50° between t = 70 s and t = −71 s. At t = −102 s, simultaneous primary and secondary shadow band modes are significant.

Fig. 3 presents a composite time-series view from the South American eclipse dataset. Gaps at selected points in the time-series plots result from transient measurement errors caused by an individual walking in front of the white sheet during the experiment. The control group used to assess statistical significance consists of data collected between C2 and C3. As C2 approaches, both the scintillation index and the primary Orientation Prominence Metric ($OPM_1$) become statistically significant. Following C3, both metrics decrease toward baseline levels. Fig. 4 displays secondary shadow band modes. At t = −102 s, the secondary Orientation Prominence Metric ($OPM_2$) becomes significant. The primary spatial band spectrogram shows an increase in power for higher spatial frequencies leading up to C2 and declining after C3. The significant primary ridge is overlaid on the spectrogram. The overall velocity of the fundamental band is around 4 m/s. Fig. 5 displays the result of image differencing at t = -102 s when primary and secondary band modes exist simultaneously and are visually apparent.

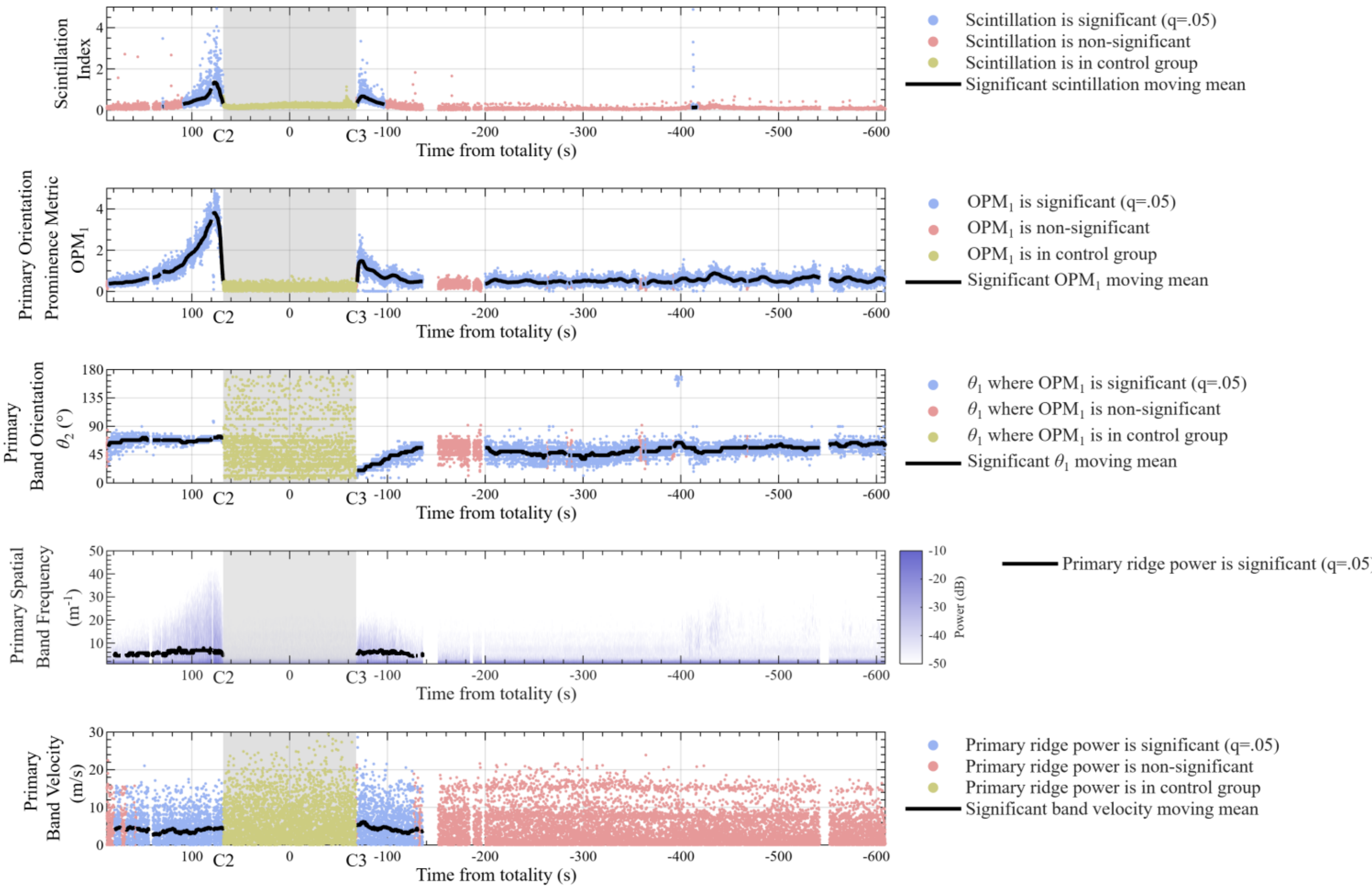

**Fig. 3. Time series data for the South America dataset.** Scintillation index and primary orientation prominence become significant during the seconds leading up to C2 and after C3. Primary spatial band spectrogram shows an increase in power for higher spatial frequencies leading up to C2 and declining after C3. The significant primary ridge is overlaid on the spectrogram. The overall velocity of the fundamental band is around 4 m/s.

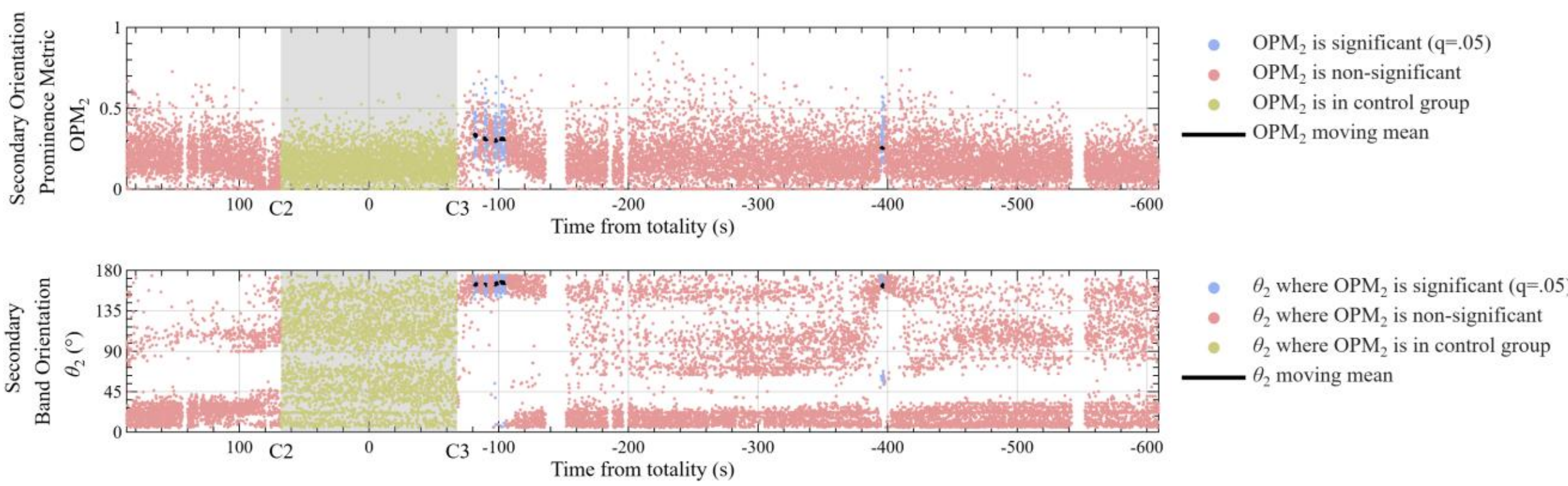

**Fig. 4. Time series data for secondary shadow band modes in the South America dataset.** At t = −102 s, secondary shadow band modes are significant at approximately 170°.

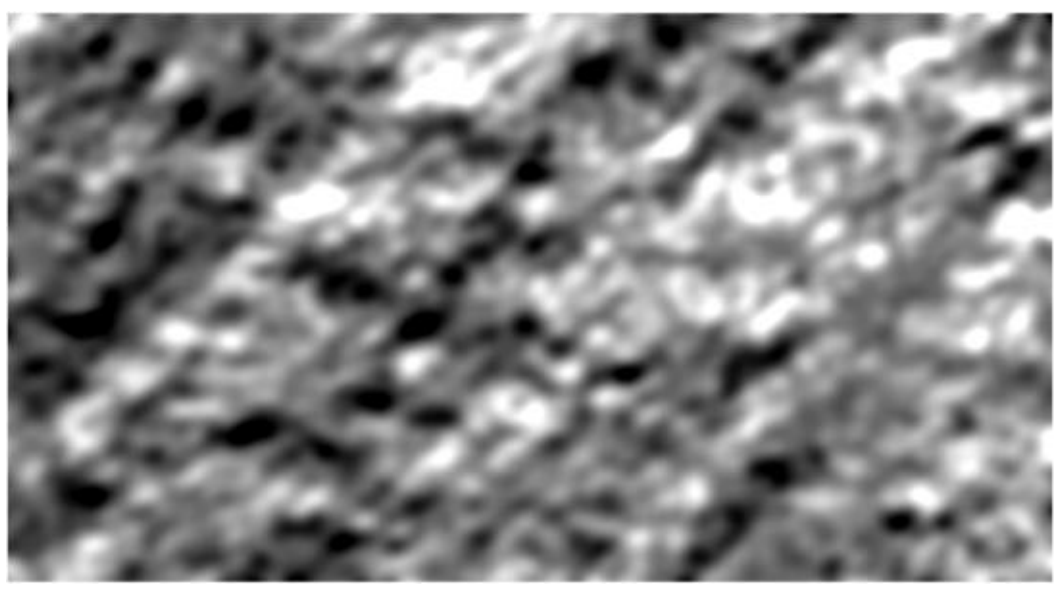

**Fig. 5. Image of simultaneous orthogonal shadow bands.** This image is the result of image differencing for the South American dataset at t=-102 s.

## Results: 2024 North America Eclipse

The second experimental dataset was recorded on April 8, 2024, during the solar eclipse that passed over North America. An iPhone 13 camera was mounted on a short stand above a white sheet measuring approximately 0.9 $m^2$. The measurement location was 45.2711° N, 72.1639° W in Magog, Canada, along the eclipse center line. The white sheet was oriented toward the northeast, with its principal axis measured at +30° relative to geographic north. Video recording was manually initiated approximately 90 seconds before C2. Fig. 6 presents the results of image differencing, ODF analysis, and band locations for the North America eclipse dataset at selected time points. At t = 140 s, no bands are present and the ODF is flat, indicating no dominant shadow band orientation. At t = 120 s and t = 110 s, the ODF exhibits a peak at approximately $\theta = 163°$, and the bands are significant relative to the control group, where $\theta = 0°$ corresponds to the positive x-axis of the transformed image and $\theta = 90°$ to the positive y-axis. At t = 100 s, after C2 has passed, the ODF returns to a flat distribution, and no primary bands are detected.

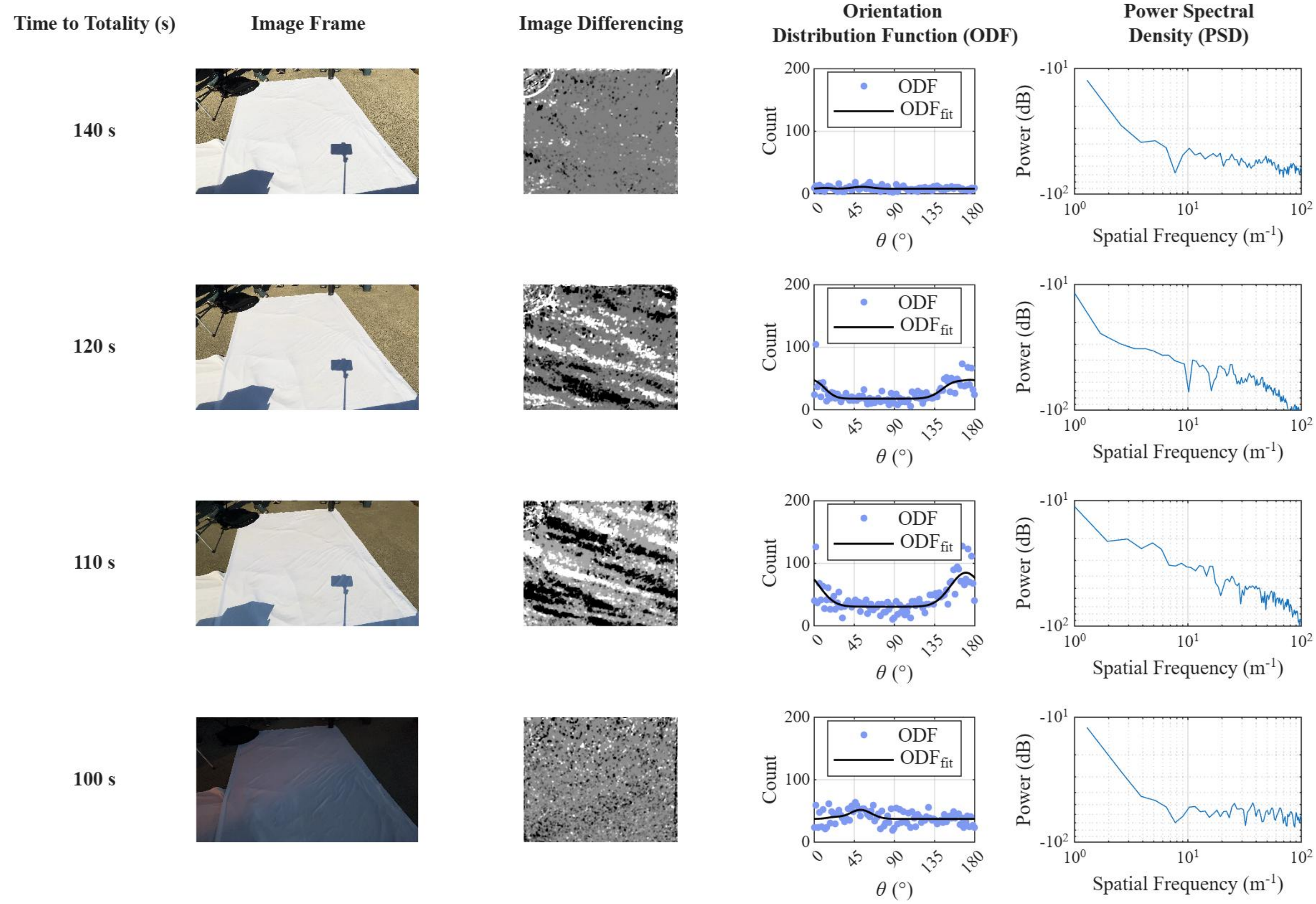

**Fig. 6. Image processing framework applied to the North American eclipse video data.** At t=120 s and 110 s, primary shadow bands at an orientation of approximately θ = 163° are significant.

Fig. 7 presents a composite time-series view of the North American eclipse dataset. Gaps at selected points in the time-series plots result from transient measurement errors caused by an individual walking in front of the white sheet during the experiment. The control group used to assess statistical significance consists of data collected after C2 and before C3, where no shadow bands are expected to be present. As C2 approaches, both the scintillation index and the primary Orientation Prominence Metric ($OPM_1$) become statistically significant. Unfortunately, no data was available after C3 because the recording stopped prematurely. A spectrogram of time-series PSD values shows a substantial increase in spatial band frequency leading up to C2. The significant primary ridge is overlaid on the spectrogram. The overall velocity of the fundamental band is around 7 m/s.

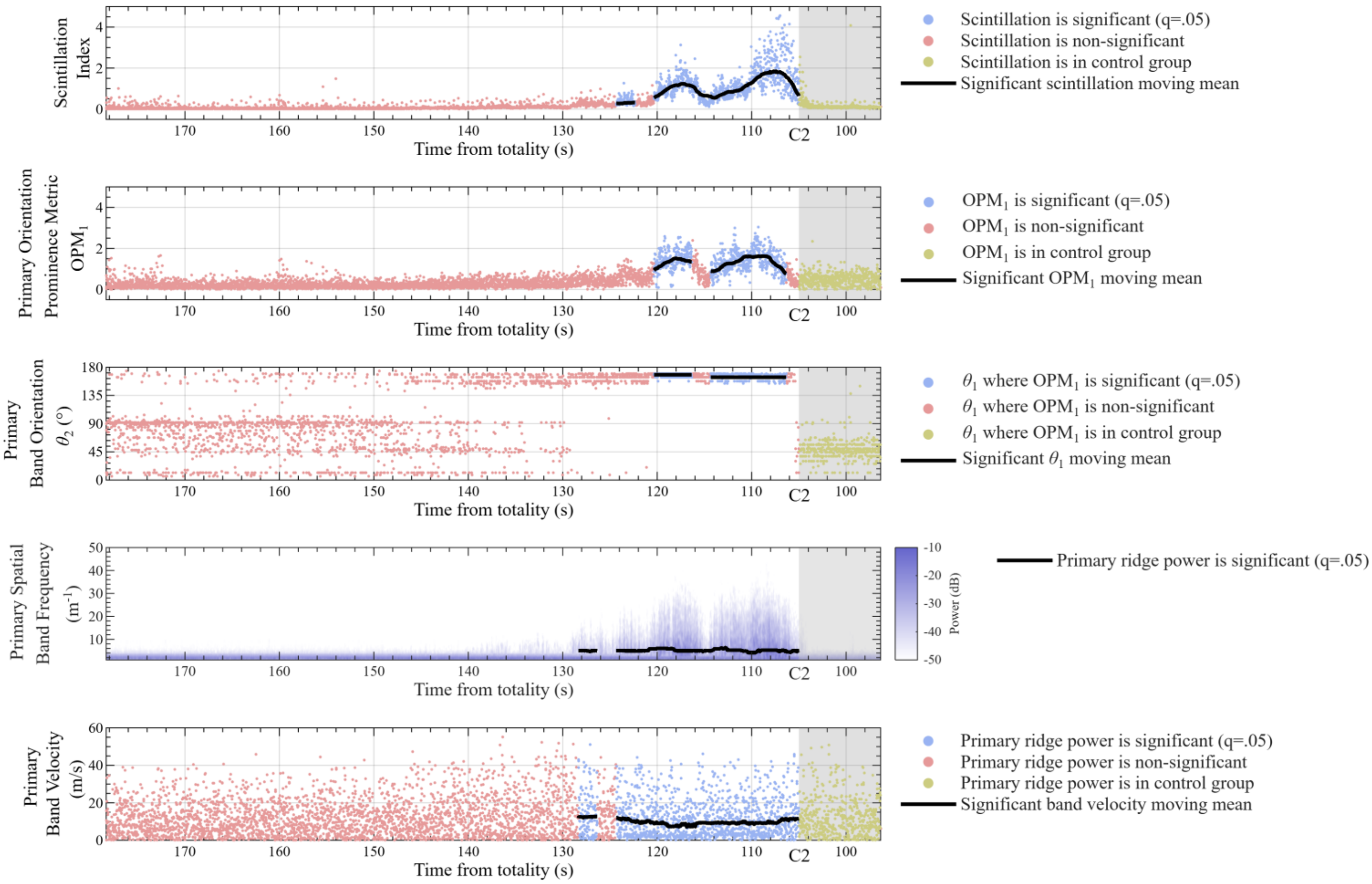

**Fig. 7. Time series data for the North America dataset.** Scintillation index and primary orientation prominence become significant during the seconds leading up to C2. The PSD spectrogram shows an increase in higher spatial frequencies leading up to C2.

## Discussion

The image-processing framework presented in this study provides a quantitative method for measuring eclipse shadow band prominence, band orientation, band separation, band velocity and power spectral density using readily obtainable ground-based video recordings. Consistent with prior research, the scintillation index increases significantly in the seconds leading up to C2 and decreases after C3 (2,3,4,5,6,7). Using an Orientation Distribution Function (ODF) and an Orientation Prominence Metric (OPM), this study quantitatively measures several subtle aspects of the eclipse shadow band phenomenon, providing greater insight into atmospheric scintillation effects.

The spatial PSD spectrogram analysis generally supports Codona's scintillation model in which higher frequencies, under appropriate atmospheric conditions, can increase in power leading up to C2 and subsequently diminish after C3. For an observer, this can create the perception that the bands are becoming closer together leading up to C2. The velocity of the shadow bands was measured between approximately 4 and 7 m/s, consistent with scintillation theory, in which band velocity is related to atmospheric wind velocity. This study found that eclipse shadow band orientation can rotate by up to 50° from a single observation location over the course of minutes. This rotation aligns with the original scintillation theory described in Codona's study when the observer is not directly on the eclipse center line. This study observed a scenario in which multiple eclipse shadow band modes can exist simultaneously, with orientations and velocities offset by up to 90°. One hypothesis is that this effect arises from multiple atmospheric layers contributing to different scintillation modes. An additional hypothesis is that these unique shadow band effects are more likely to occur during eclipses near sunrise or sunset, when light travels through a longer column of atmosphere. In conclusion, this study demonstrates that consumer-grade video recordings, when analyzed using the proposed methodology, can reveal detailed and dynamic shadow band behavior. These results further support the scintillation theory and enable improved modeling and future observational studies.

**Limitations**

This study is based on two eclipse datasets, which limits the statistical generality of the findings. The analysis is sensitive to environmental disturbances, including occlusions and surface artifacts, which required manual mitigation. Additionally, no direct atmospheric measurements (e.g., wind profiles or turbulence structure) were available, limiting the ability to correlate observed band behavior with physical atmospheric conditions. The framework also depends on camera geometry and preprocessing parameter choices, which may introduce variability across datasets. Future work incorporating additional observations and synchronized atmospheric measurements would improve robustness and physical interpretation.

## Acknowledgments

This manuscript was reviewed and edited for grammar, clarity, and style with the assistance of OpenAI's ChatGPT language model. Perplexity AI was used to check references. The MATLAB Copilot AI model was used to suggest different image processing approaches to the author. The author thanks Dr. Thierry Pauwels of the Royal Observatory of Belgium for providing experimental data collected during the 2019 solar eclipse in South America for use in this research.